\documentclass[pre,twocolumn,showpacs,aps,floatfix,superscriptaddress]{revtex4}

\usepackage[english]{babel}
\usepackage{amssymb,amsfonts,amsmath}
\usepackage[pdftex]{graphicx}
\usepackage{latexsym}
\usepackage{multirow}
\usepackage{color}

\newcommand{\mP}{\mathcal{P}}
\newcommand{\mT}{\mathcal{T}}

\newcommand{\mR}{\mathcal{R}}

\newcommand{\mV}{\mathcal{V}}
\newcommand{\mG}{\mathcal{G}}

\begin{document}

\title{Combinatorial approach to Modularity}

\author{Filippo Radicchi}\affiliation{Complex Networks Lagrange Laboratory (CNLL), ISI Foundation, Turin, Italy}

\author{Andrea Lancichinetti}\affiliation{Complex Networks Lagrange Laboratory (CNLL), ISI Foundation, Turin, Italy}\affiliation{Physics Department, Politecnico di Torino, Turin, Italy}

\author{Jos\'e J. Ramasco}\affiliation{Complex Networks Lagrange Laboratory (CNLL), ISI Foundation, Turin, Italy}

\begin{abstract}
Communities are clusters of nodes with a higher than average density of internal connections. Their detection is of great relevance to better understand the structure and hierarchies present in a network. Modularity has become a standard tool in the area of community detection, providing at the same time a way to evaluate partitions and, by maximizing it, a method to find communities. In this work, we study the modularity from a combinatorial point of view. Our analysis (as the modularity definition) relies on the use of the configurational model, a technique that given a graph produces a series of randomized copies keeping the degree sequence invariant. We develop an approach that enumerates the null model partitions and can be used to calculate
the probability distribution function of the modularity. 
Our theory allows for a deep inquiry of
several interesting features characterizing modularity such as its
resolution limit and the statistics of the partitions that maximize it. Additionally, the study of the probability of extremes of the modularity in the random graph partitions opens the way for a definition
of the statistical significance of network partitions.
\end{abstract}


\maketitle

\section{Introduction}

Graphs are used as mathematical representations of complex systems. Examples can be found in biology, technology, social and information sciences~\cite{albert02,newman03,romu_alex04}. Real world networks show several non trivial topological features, among which one of the most fascinating is the organization of their nodes in local clusters or modules known as communities. Communities are groups of nodes with a high level of internal and low level of external connectivity. They are subgraphs relatively isolated from the rest of the network and are expected to correspond to groups of elements sharing common features and/or playing similar roles within the original system. The last few years have witnessed an increasing interest in defining and identifying communities~\cite{girvan02,zhou03,newman04,radicchi04,reichardt04,palla05,guimera05,arenas06,rosvall08} (see~\cite{santo10} for a recent review). Different methods have been proposed from topological considerations~\cite{newman04,radicchi04,palla05} to the study of the influence that communities have in the properties of dynamical processes running on the network such as random walks diffusion~\cite{zhou03,rosvall08} or the Potts model~\cite{reichardt04}. 
 
A major role in this context is played by the modularity function $Q$ introduced by Newman and Girvan~\cite{newman04}.
The modularity is a quality measure aimed at quantifying the relevance of the community structure in a network partition. It is defined as 
\begin{equation}
Q_C = \frac{1}{M} \sum_{\varphi =1}^{C} \left(e_{\varphi,\varphi} - \langle e_{\varphi,\varphi} \rangle \right) ,
\end{equation}
where $M$ is the total number of links in the network, the sum runs over the $C$ communities of the partition, $e_{\varphi,\varphi}$ stands for the number of internal links in the community $\varphi$, and $\langle e_{\varphi,\varphi} \rangle$ is the expected value of this quantity in a  random null model (typically, the configurational model). The modularity corresponds thus to the comparison between the actual number of internal links of the modules and the number they would have in a random null model. The partition with maximal $Q$ is then considered  the best and most significant division of the network in communities~\cite{newman04}. The search for such optimal partition is in general a great challenge since it was proved to be a NP-complete hard problem~\cite{brandes08}. Many heuristics relying on different approaches have been introduced to approximate the optimal partition: Some based on cluster hierarchical division or aggregation methods~\cite{newman04,newman04b,clauset04,danon06,wakita07,arenas07,blondel08,schuetz08,mei09}, on simulated annealing~\cite{guimera05,massen06}, spectral methods~\cite{newman06,leicht08,sun09}, genetic algorithms~\cite{bingol07} or extremal optimization~\cite{duch05} to mention a few. Still modularity maximization as a procedure for community detection is not free from shadows. It was shown that the modularity suffers from resolution limits~\cite{fortunato07,kumpula07}, not being able to discern the quality of modules smaller than a certain size ($\sqrt{M}$). Also optimized partitions even in random graphs have non zero modularity~\cite{guimera04}, posing the question of the significance of a partition. And, finally, the huge number of degenerate local maxima of $Q$ in common examples can practically prevent the finding of the real optimal partition~\cite{good09}.

In this paper, we choose a different route to study the modularity function, trying to shed some additional light on its limits and intrinsic properties. We develop a combinatorial method to estimate the distribution of modularity values in the partitions of the configurational model~ \cite{molloy98,boguna04,catanzaro05}. Our approach leads us to write explicit formulas for the modularity distribution and to analyze in details the characteristics of this function. We focus our attention on the resolution limit of modularity~\cite{fortunato07} showing that, even in the case of random networks, modularity prefers to merge small groups into larger ones. 
We also focus on the evaluation of the statistical significance of communities, basing our estimates on the probability associated to modularity in the configurational model and extending  previous results on the topic~\cite{guimera04,lancichinetti10}. 

The paper is organized as follows. In section~\ref{sec:model}, we introduce the configurational model (i.e., the {\it null model} of modularity) and propose a combinatorial approach for the study of its networks' partitions.  In particular, subsection~\ref{sec:conf_model} is devoted to the description of the model, while subsection~\ref{sec:conf_model_comm} deals with the theory of communities in the configurational model. In subsection~\ref{sec:int_conn}, we show how to estimate the number of internal connections of a community. From section~\ref{sec:mod}, we start with the analysis of the modularity function in the configurational model. We show exact expressions for the probability distribution function of modularity and analyze its main features. In section~\ref{sec:stat}, we focus on the statistics of the maximal modularity in the configurational model and propose a simple, but efficient way for the determination of the statistical significance of partitions in networks. In section~\ref{sec:dirntw}, we extend our whole theory to the case of directed and bipartite networks. We draw our final comments and considerations in section~\ref{sec:concl}.

\section{Statistical Model}
\label{sec:model}       

The configurational model is a prototypical algorithm for the generation of uncorrelated networks with prescribed number of nodes and of node connections (degree). The procedure for the random networks  construction was originally introduced by Molloy and Reed in Ref.~\cite{molloy98}. This model has been the subject of many research papers along the last decade. Typical properties observed in real networks are generally tested against the model graphs in order to asses whether they are effectively genuine or just induced by the constraints to which the network is subjected as keeping a degree sequence invariant. Examples range from the simple determination of degree-degree correlations~\cite{newman02} to clustering~\cite{newman09}. Community structure, which can be seen as a correlation between connections at a local level, is (must be) also tested against a null model. The modularity function, which has become the standard tool in community detection, is defined using the configurational model as null model~\cite{newman04}. Modularity in fact compares the number of connections between nodes of the same module with the one expected on average in the configurational model, i.e., for random networks with the same set of vertices and node degrees as the given graph. Before going further, it is worth stressing that other, more or less restrictive, null models can be also employed in defining the modularity function~\cite{massen05,gaertler07,nicosia09}. We chose the configurational model as a paradigmatic example for our analysis essentially due to its simplicity, to the fact that it was the original null model in the definition of $Q$ and that it keeps being the most extensively used.

In the next subsections, we study in details the configurational model. We propose a combinatorial approach for the enumeration of all possible network partitions belonging to the ensemble generated by the model and formulate exact expressions for the probability of the number of internal connections of their modules. The whole theory represents therefore a combinatorial approach to the configurational model with explicit application to modularity.

\subsection{The configurational model}
\label{sec:conf_model}

\begin{figure}[b]
\includegraphics[width=0.45\textwidth]{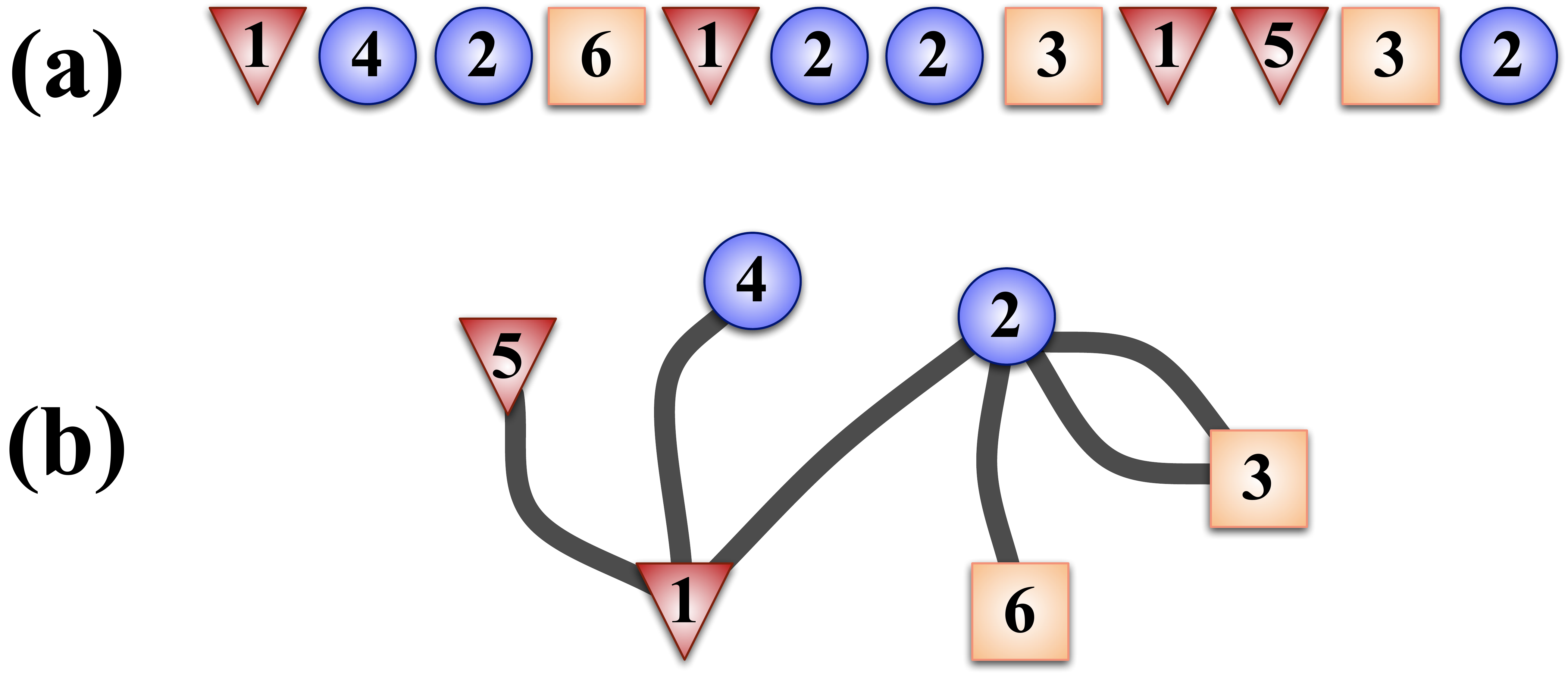}
\caption{(Color online) A simple network generated according to the configurational model.
The network is composed of $N=6$ nodes and $M=6$ edges. The degree sequence is 
$\{k_i\}= \{k_1=3, k_2=4, k_3=2, k_4=1, k_5=1, k_6=1\}$. (a) A sequence of
node labels is generated and (b) according to it connections are drawn in the network.
If node labels are replaced by community labels  (in this case, $\sigma_1 = \sigma_5 =\bigtriangledown, \sigma_2=\sigma_4=\bigcirc , \sigma_3=\sigma_6=\Box$),  the network in (b) can be seen as a graph between $C=3$ communities with degree sequence
$\{d_\alpha\}=\{d_\bigtriangledown=4, d_\bigcirc=5, d_\Box=3\}$. In this particular case,
the measured values of intra- and inter-community connections are: $\{e_{\alpha,\alpha}, e_{\alpha,\beta}\} =  \{e_{\bigtriangledown, \bigtriangledown}=1, e_{\bigcirc, \bigcirc}= 0, e_{\Box, \Box}=0, e_{\bigtriangledown, \bigcirc}=2, e_{\bigtriangledown, \Box}=0, e_{\bigcirc, \Box}=3 \}$.}
\label{fig1}
\end{figure}

The basic ingredients of the configurational model are the number of nodes and the degree sequence of the network nodes. Consider therefore a network composed of $N$ nodes and denote the degree of the $j$-th node by $k_j$. The full degree sequence is then the set $\{k_i\}=\{k_1, k_2, \ldots, k_N\}$. The procedure to construct the networks is very simple: each node $j$ is connected to other $k_j$ randomly chosen nodes but always satisfying the constraints imposed by keeping the entire degree sequence constant. We consider first the case of undirected networks. For this class of networks, the sum of all degrees should be an even number and we can thus write 
\begin{equation}
\sum_{j=1}^N k_j = 2 M \,\,\,.
\label{eq:constr_1}
\end{equation}
The generation mechanism of the configurational model can be formulated in an alternative manner (see Figure~\ref{fig1}):  ({\it i}) Randomly fill a list composed of $2M$ entries with node labels ranging from $1$ to $N$, where the number of appearances of each label is equal to the respective node degree;  ({\it ii}) Draw a connection between each pair of nodes whose labels appear at positions $p_{2k-1}$ and $p_{2k}$ for each $k=1, 2, \ldots, M$. It is clear that, in the case of this construction procedure, multiple connections and self-loops are not avoided. Their presence however can be considered negligible under certain realistic assumptions~\cite{catanzaro05}, in simple words that no node concentrates a significant fraction of the network connections~\cite{note}.

The  construction procedure just introduced is the most common technique to build the graphs of the configurational model. Note that it samples homogeneously out of the set of all possible sequences of node labels, not out of the set of all possible graphs with given degree sequence. The reason is that the same graph may be represented by different sequences of node labels and its multiplicity may vary as a function of several factors (i.e., number of self-loops,
multiple connections, etc.).  The total number of possible  sequences of node labels with prescribed degree sequence $\{k_i\}$ is simply given by
\begin{equation}
\mT_N \left(\{k_i\}\right) = {2M \choose k_1, k_2, \ldots, k_N} = \frac{\left(2M\right)!}{k_1!\,k_2!\,\cdots\,k_N!} \,\,\, .
\label{eq:tot_nl}
\end{equation}
The term on the right of Eq.~(\ref{eq:tot_nl}) is a multinomial coefficient and counts the total number of ways of organizing $N$ node labels with multiplicities $\{k_i\}$ subjected to the constraint of Eq.~(\ref{eq:constr_1}).

\subsection{Communities in the configurational model}
\label{sec:conf_model_comm}

Consider now a partition of the network in $C$ groups or communities. For partition we mean a division of nodes in several non overlapping node groups. The degree $d_\varphi$ of the group $\varphi$ (where $\varphi$ can be $1,\ldots,C$) is given by the sum of the degrees of all nodes belonging to it
\begin{equation}
d_\varphi = \sum_{j \in \varphi}  k_j \,\,\,,
\label{eq:degree_comm}
\end{equation}
The network between communities in the configurational model is equivalent to a configurational model composed of $C$ "super nodes", one per group, with degree sequence $\{d_\alpha\} = \{d_1, d_2, \ldots, d_C\}$ (see Figure~\ref{fig1}). Similarly to the argument leading to Eq.~(\ref{eq:tot_nl}), also in this case the total number of sequences of communities labels can be written as 
\begin{equation}
\mT_C \left(\{d_\alpha\}\right)={2M \choose d_1,\,d_2,\,\ldots\,,d_C}=\frac{\left(2M\right)!}{d_1!\,d_2!\,\cdots\,d_C!}\,\,\,.
\label{eq:tot_nl2}
\end{equation}
If we refer as $e_{\varphi,\theta}$ to the number of edges present between
the $\varphi$-th and the $\theta$-th community since the network is undirected we have for symmetry that $e_{\varphi,\theta}=e_{\theta,\varphi}$, for any $\varphi$ and $\theta$. The links intra-community are completed by the internal group links, denoted as $e_{\varphi,\varphi}$ for each group $\varphi$. By definition, these quantities should obey the $C$ relations
\begin{equation}
d_\varphi = e_{\varphi,\varphi} + \sum_{\theta=1}^C e_{\varphi,\theta} \;\; , \;\; \forall \, \varphi=1, \ldots, C \,\,,
\label{eq:constr_2}
\end{equation}  
because the degree of the  $\varphi$-th community is equal to the sum of all edges having only one end in $\varphi$ plus twice the number of edges having both ends in the group. Fixed a particular set of values for intra- and inter-community edges, namely $\{e_{\alpha,\alpha}, e_{\alpha,\beta}\} = \{e_{1,1}, e_{2,2}, \ldots, e_{C,C}, e_{1,2}, e_{1,3}, \ldots, e_{1,C}, \ldots, e_{C-1,C}\}$, the total number of sequences of community labels that satisfy these requirements are
\begin{align} 
\label{eq:tot_nc}
\mR_C \left( \{e_{\alpha,\alpha}, e_{\alpha,\beta}\} \right) = & \,\; \\
M!\, \prod_{\varphi=1}^C \frac{1}{e_{\varphi,\varphi}!}  & 2^{\sum_{\varphi=1}^{C-1} \sum_{\theta=\varphi+1}^C e_{\varphi,\theta}} \, \prod_{\varphi=1}^{C-1} \, \prod_{\theta=\varphi+1}^{C} \frac{1}{e_{\varphi,\theta}!} \,\,\, . \nonumber
\end{align}
Eq.~(\ref{eq:tot_nc}) states that the number of sequences of community labels, 
with given intra- and inter-community edges $\{e_{\alpha,\alpha}, e_{\alpha,\beta}\}$, can be obtained as the product of three factors: ({\it i}) $M!$, the number of permutations of the $M$ edges; ({\it ii}) $\prod_{\varphi=1}^C \frac{1}{e_{\varphi,\varphi}!}$, the inverse of the different number of times to list all the intra-community edges; and ({\it iii}) $ \prod_{\varphi=1}^{C-1}\,\prod_{\theta=\varphi+1}^{C} \frac{1}{e_{\varphi,\theta}!}$, the inverse of the total number of ways to arrange all the inter-community edges, where in particular the factor $2^{\sum_{\varphi=1}^{C-1} \sum_{\theta=\varphi+1}^C e_{\varphi,\theta}}$ is needed due to the fact that the presence of an inter-community edge is independent of the order in which the community labels appear on the list (i.e., $e_{\varphi,\theta} \equiv e_{\theta,\varphi}$, for any $\varphi$ and $\theta$). The probability therefore to observe a particular sequence of label communities with certain set of values  $\{e_{\alpha,\alpha}, e_{\alpha,\beta}\}$  is given by the ratio between the quantities defined in Eqs.~(\ref{eq:tot_nc}) and~(\ref{eq:tot_nl2}),
\begin{equation}
\label{eq:prob_tot}
\mP_C \left( \{e_{\alpha,\alpha}, e_{\alpha,\beta}\} \right) = \frac{\mR_C \left( \{e_{\alpha,\alpha}, e_{\alpha,\beta}\} \right)}{\mT_C \left(\{d_\alpha\}\right)} .
\end{equation}

\subsection{Internal connectivity of communities}
\label{sec:int_conn}

In the case of communities, we are not generally interested in the whole set $\{e_{\alpha,\alpha}, e_{\alpha,\beta}\}$ for the intra- and inter-community edges, but  only in the set of possible sequences with given intra-community
edge sequence $\{e_{\alpha,\alpha}\}$. This basically amounts to calculating the marginal distribution of the probability in Eq.~(\ref{eq:prob_tot}) by summing
over all the possible configurations of the inter-community edges $\{e_{\alpha,\beta}\}$
\begin{equation}
\mP_C \left( \{e_{\alpha,\alpha}\} \right) = \frac{1}{ \mT_C\left( \{ d_\alpha \} \right)} \, \sum_{\{e_{\alpha,\beta}\}}  \mR_C \left( \{e_{\alpha,\alpha}, e_{\alpha,\beta}\} \right)\, \,\,,
\label{eq:prob}
\end{equation}
where in the sum the inter-community edges $\{e_{\alpha,\beta}\}$ are subjected to the constraints of Eqs.~(\ref{eq:constr_2}).

$\mP_C \left( \{e_{\alpha,\alpha}\} \right)$ is the probability that groups of nodes, with degrees specified by $\{d_\alpha\}$, have internal connections equal to the sequence $\{e_{\alpha,\alpha}\}$ in the hypothesis that connections have been drawn according to the configurational model rules. The distribution $\mP_C \left( \{e_{\alpha,\alpha}\} \right)$ can be easily obtained for $C=2$ and $C=3$. In these cases, the inter-community edges $\{e_{\alpha,\beta}\}$ are completely determined by the constraints of Eqs.~(\ref{eq:constr_2}) given the number of intra-community edges $\{e_{\alpha,\alpha}\}$, hence no sum is actually required. For example, for $C=2$ Eq.~(\ref{eq:prob}) becomes
\begin{align}
\label{eq:prob2}
\mP_2\left(\left\{e_{1,1}\right\}\right) =  & \,\;  \\ 
\, & \frac{M!}{\left( 2M\right)!} \,  \frac{d_1!\,\left(2M-d_1\right)! \; 2^{d_1-2e_{1,1}}}{e_{1,1}!\,\left(M-d_1+e_{1,1}\right)! \,\left(d_1-2e_{1,1}\right)!} \, ,  \nonumber 
\end{align}
given that from Eqs.~(\ref{eq:constr_1}) and~(\ref{eq:constr_2}) we have
$e_{1,2}=d_1-2e_{1,1}$ and $e_{2,2}=M-d_1+e_{1,1}$. Notice that
$\mP_2\left(\left\{e_{1,1}\right\}\right)$ depends only on $e_{1,1}$, since $e_{2,2}$ is fixed
for any value of $e_{1,1}$ and {\it viceversa}. Interestingly, the distribution  of Eq.~(\ref{eq:prob2}) has been also found as the solution of a completely different problem in survival analysis where is known as the
{\it  Univariate Twins Distribution} and has applications also to the study of the genetic variability of neutral alleles in a population~\cite{zelter}.

For $C=3$, the calculations are a little more cumbersome but we obtain
\begin{align}
\label{eq:prob3}
\mP_3\left(\left\{e_{1,1}, e_{2,2}, e_{3,3}\right\}\right) = &\,\; \\
 \frac{M!}{\left( 2M\right)!}\, 2^{M-M{int}} & \prod_{\varphi=1}^3 \frac{d_\varphi!}{e_{\varphi,\varphi}!\; \left( M - M_{int}-d_\varphi+2e_{\varphi,\varphi} \right)!} \,, \nonumber
\end{align}
where $M_{int}=\sum_{\varphi=1}^3 \, e_{\varphi,\varphi}$ is the total 
number of intra-community edges.

The general case (i.e., arbitrary number of groups $C$) includes a sum over all the possible configurations of the inter-groups connections. This turns the calculation of $\mP_C\left(\left\{e_{\alpha,\alpha}\right\}\right)$ quite hard, in fact we were not able to  find an analytical closed form for it. This problem is similar to those appearing in the enumeration of contingency tables (whose most celebrated examples are the {\it latin} and {\it magic squares}) and represents still an open problem in combinatorics~\cite{good76,jucys77,metha83}. It is still possible to numerically determine the sum with a computational time growing as $M^{C^2}$ [the number of free indices in the sum of Eq.~(\ref{eq:prob}) is $C^2/2-3C/2$]. Another possibility is to relax the constraints of Eqs.~(\ref{eq:constr_2}) considering the groups as independent of each other. This ``pair approximation'' yields 
\begin{equation}
\mP_C \left( \{e_{\alpha,\alpha}\} \right) \simeq \tilde{\mP}_C \left( \{e_{\alpha,\alpha}\} \right) = \prod_{\varphi=1}^C \; \mP_2\left(\left\{e_{\varphi,\varphi}\right\}\right) \;\;,
\label{eq:prob_approx}
\end{equation}
which stands for the product of $C$ independent bipartitions, each of them weighted by the probability $\mP_2\left(\left\{e_{\varphi,\varphi}\right\}\right)$ of Equation~(\ref{eq:prob2}), where the constraints are now simply 
$2e_{\varphi,\varphi} \leq d_\varphi$, $\forall \varphi=1, \ldots, C$. Due to the reduced calculation burden, this approximation can be helpful in some cases in which a fast evaluation of $\mP_C \left( \{e_{\alpha,\alpha}\} \right)$ is needed. We expect it to work better when the number of communities $C$ is larger.

\section{Modularity function}
\label{sec:mod} 

\subsection{Modularity distribution in the configurational model}

Up to now we have introduced a formalism which allows to compute, given $C$ groups of nodes and their degree sequence $\{d_\alpha\}$, the probability distribution function that such groups have a set $\{e_{\alpha,\alpha}\}$ of internal connections under the hypothesis that the network is generated according to the configurational model algorithm. As explained before, the modularity function $Q_C$ of a partition in $C$ groups with degree
sequence $\{d_\alpha\}$ and internal connectivities $\{e_{\alpha,\alpha}\}$ is defined as
\begin{equation}
Q_C = \frac{1}{M} \, \sum_{\varphi=1}^C  \left(e_{\varphi,\varphi} - \langle e_{\varphi,\varphi}  \rangle \right) = \frac{M_{int} - \mV_C\left(\{d_\alpha \} \right)}{M} \;,
\label{eq:mod}
\end{equation}
where $\mV_C\left(\{d_\alpha \} \right)= \sum_{\varphi=1}^C \langle e_{\varphi,\varphi}  \rangle$
represents the sum of the expected internal connectivities over all modules
and is determined by the degree sequence of the modules $\{d_\alpha \}$.
The average value of the intra-community edges of the
module $\varphi$ can be obtained by marginalizing the general distribution 
$\mP_C \left( \{e_{\alpha,\alpha}\} \right)$
of Eq.~(\ref{eq:prob}) and turns out to be
\begin{equation}
\langle e_{\varphi,\varphi}  \rangle = \frac{d_\varphi \left(d_\varphi-1\right) }{2 \left(2M-1\right)} \;.
\label{eq:av_mod}
\end{equation}
Notice that this average value is slightly different from the one used in the original formulation of the modularity, {i.e., $\langle e_{\varphi,\varphi}  \rangle = \left(d_\varphi\right)^2/\left(4M\right)$, which is a rougher approximation to the value expected in the configurational model. The probability of the modularity function to have a value $Q$ for the networks of the null model ensemble can be then calculated 
as
\begin{equation}
\mP_C \left(Q \right) = \sum_{\{e_{\alpha,\alpha}\}} \, \mP_C \left( \{e_{\alpha,\alpha}\} \right) \, \delta \! \left[ Q_C\left( \{e_{\alpha,\alpha}\} \right) - Q \right] \;.
\label{eq:prob_mod}
\end{equation}
Note that the term $\delta \!\left[ Q_C\left( \{e_{\alpha,\alpha}\} \right) - Q \right]$ adds to Eqs.~(\ref{eq:constr_2}) the new constraint  $M_{int} = M \,Q + \mV_C\left(\{d_\alpha \} \right)$. For instance, this implies that for $C=2$ and $C=3$ the distribution of the modularity in the configurational model
can be obtained by modifying accordingly Eqs.~(\ref{eq:prob2}) and~(\ref{eq:prob3}).

\subsection{Properties of $Q_C$ and $\mP_C(Q)$}

\begin{figure*}
\includegraphics[width=0.3\textwidth]{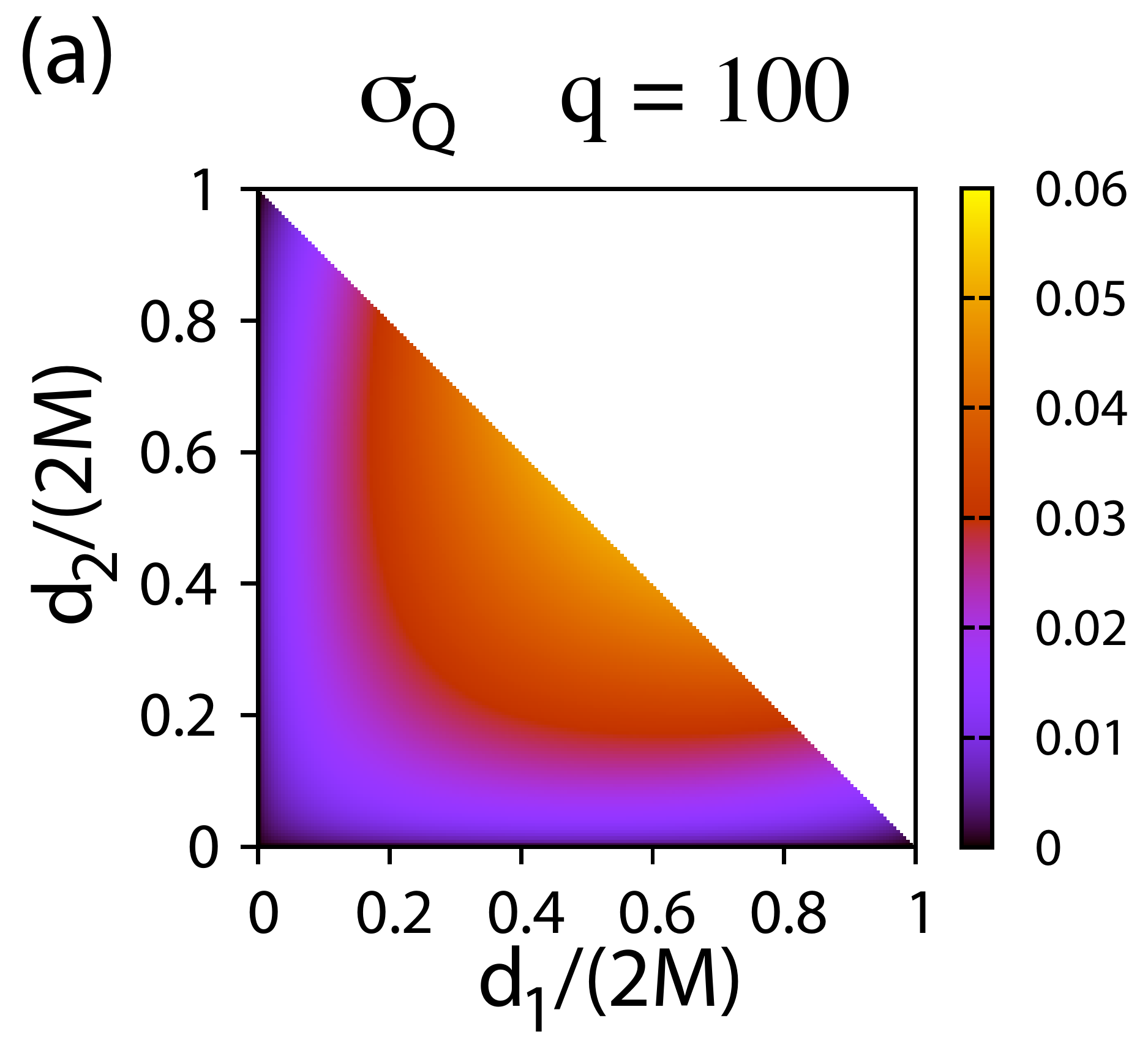}
\quad
\includegraphics[width=0.3\textwidth]{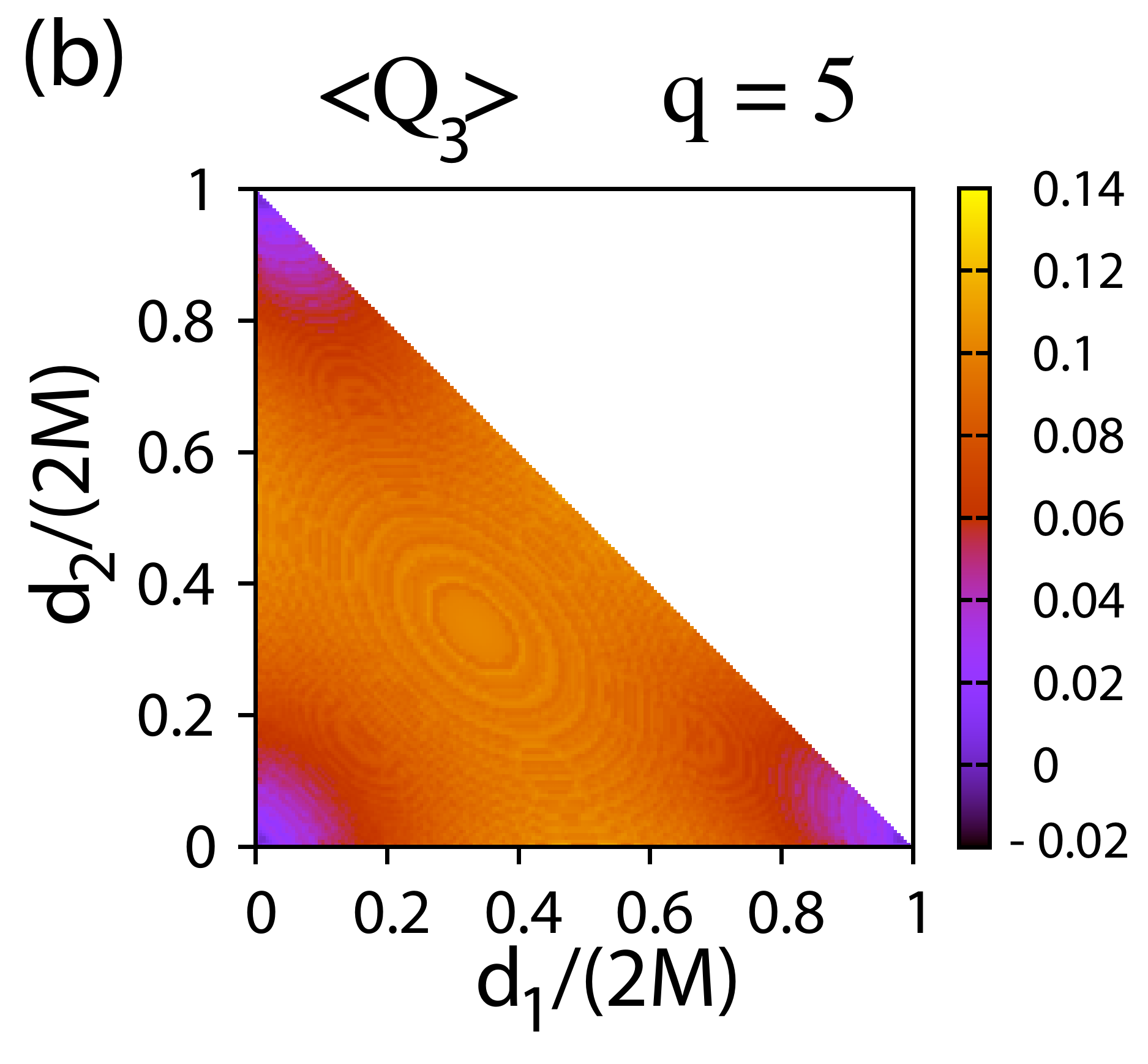}
\quad
\includegraphics[width=0.3\textwidth]{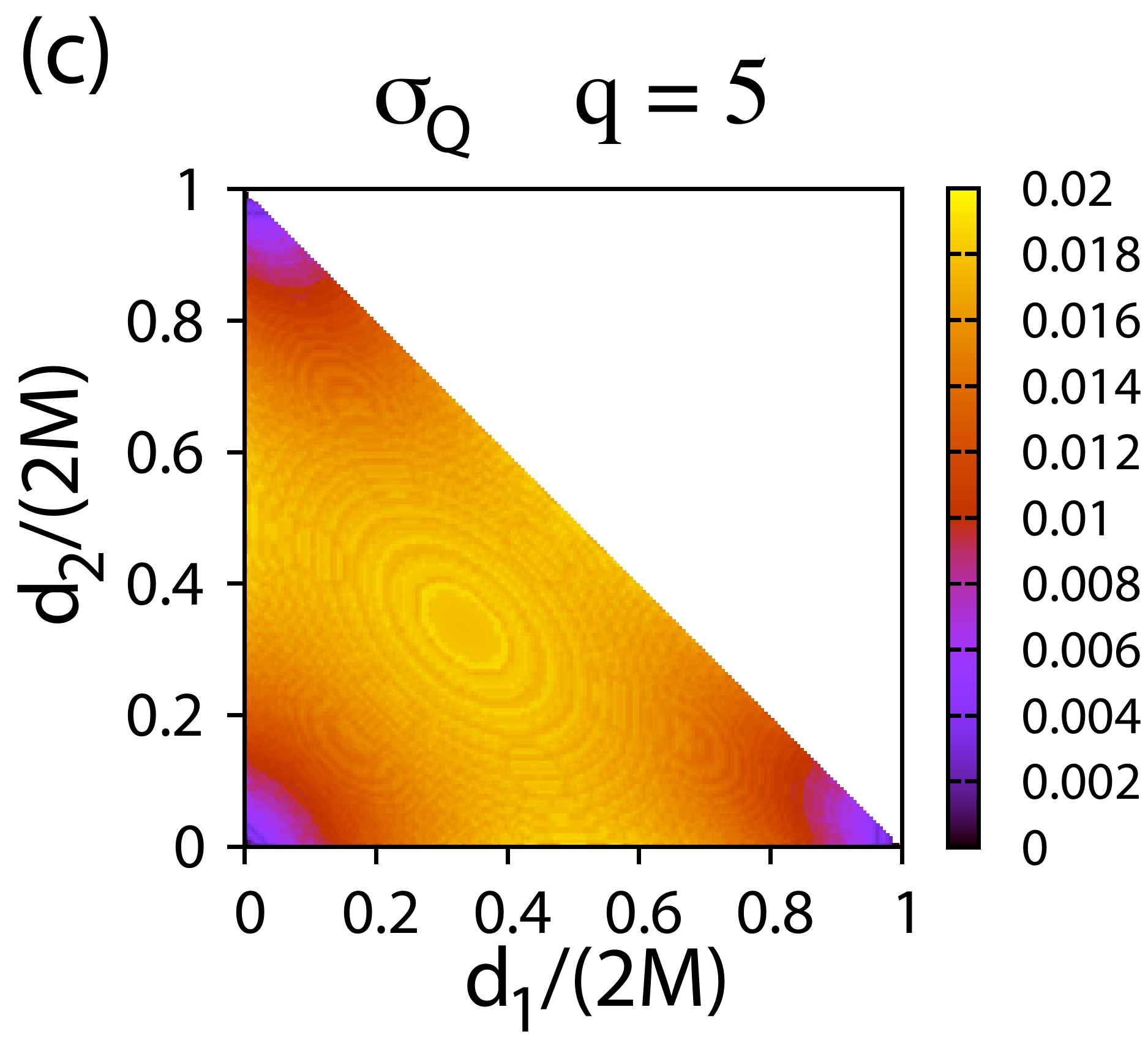}
\caption{(Color online) Fixed the partitions corresponding to top
$q\%$ of modularity, we compute the average and standard deviation, over this ensemble,
 of the modularity $Q_3$  as a function of the relative degrees 
of the groups [i.e., $d_1/\left(2M\right)$ and $d_2/\left(2M\right)$]. For $q=100$,
the average value (not shown) is zero for every value of $d_1/\left(2M\right)$ and $d_2/\left(2M\right)$.
On the other end, the standard deviation (panel a) tends to be small when  the degree 
of one of the communities is
small and grows as the communities become similar in their degrees. For $q=5$ (panel b and c),
both average value and standard deviation grows as the partition becomes more homogeneous. Here we set $M=100$.}
\label{fig:var_100}
\end{figure*}

We illustrate now some characteristics of $Q_C$ and its distribution  $\mP_C(Q)$ in the null model with a few examples simple enough to admit an analytic or semi-analytic treatment. The interest in the use of modularity is generally focused on the search of the partition with the maximum $Q_C$. This search, as has been discussed, is a hard problem~\cite{brandes08}, mainly due to the huge amount of almost degenerate local maxima in the modularity landscape~\cite{good09}. Such abundance of local maxima has been even found when the modularity optimization is applied to the random networks generated with the configurational model. With our formalism we are not able to judge whether a partition is a local maximum in $Q_C$ landscape, but we can already evidence the problem of the abundance of local structure by considering our results from a more restricted point of view. For the first of our examples, we choose to split the null model networks in three groups, a case for which we can obtain analytical solutions. We compute the average value ($\langle Q_3 \rangle$) and the standard deviation ($\sigma_Q$) of $\mP_3\left(Q\right)$ as a function of the relative degree of the communities [i.e., $d_1/\left(2M\right)$ and $d_2/\left(2M\right)$]. These quantities are calculated only over
the partitions corresponding to the top $q\%$ instances of the 
modularity. For $q=100$, $\langle Q_3 \rangle = 0$ everywhere, as expected since the expected modularity in the null model is zero,  while the standard deviation
exhibits a regular behavior. The results for $ \sigma_Q$ can be seen in the panel (a) of Figure~\ref{fig:var_100}. Then to approximate the local maxima of $Q_C$, we restrict the calculations to only the top $q = 5\%$ instances of the modularity distribution. Recall that we are doing this analytically so the analysis precision does not suffer for concentrating in extreme values. In the panel (b) of Figure~\ref{fig:var_100}, one can observe how the average
is not longer null and varies consistently from the region of imbalanced partitions [i.e., $d_\varphi/\left(2M\right) \simeq 0$ for one the $\varphi$ group] to the zone of homogeneous partitions [i.e., $d_\varphi/\left(2M\right) \simeq 1/3$ for all $\varphi$]. There is a wide region in which large changes of $d_1/\left(2M\right)$ and $d_2/\left(2M\right)$ do not produce important variations in the average value. At the same time, it is possible to observe a fine structure pointing to a rich local landscape geometry for $Q_3$. This result is just indicative since the projection of the partitions space in a plane with only two parameters [$d_1/\left(2M\right)$ and $d_2/\left(2M\right)$] is too gross. See for instance~\cite{good09} for a more systematic method to do such projection. The standard deviation of the top $5\%$ modularity instances, (panel c of the Figure~\ref{fig:var_100}), continues to be large for homogeneous partitions and decreases as the partition becomes more imbalanced following similar patterns as $\langle Q_3\rangle$.

We consider next another interesting application related to the so-called
{\it resolution limit} of the modularity function~\cite{fortunato07,kumpula07}. We analyze all the possible divisions in $C=3$ groups [as before monitored as a function of the relative degree of two groups, $d_1/\left(2M\right)$ and $d_2/\left(2M\right)$) and calculate the modularity $Q_3$. Fixed $d_1/\left(2M\right)$ and $d_2/\left(2M\right)$ [and $d_3/\left(2M\right)$], we calculate also $Q_2$ which is the modularity of the partition with groups $1$ and $2$ merged together.
The quantity $Q_3 - Q_2$ is then measured and its average value and standard deviation over all partitions corresponding to the top $q\%$ values of $Q_3$ is estimated. Note that if  $Q_2 > Q_3$ according to modularity optimization it would be more convenient to merge both communities. When all the partitions are considered (i.e., $q=100$) the average is always zero and the standard deviation (see Figure~\ref{fig:var_100a}a)
shows a regular pattern with maximum at $d_1/\left(2M\right) = d_2/\left(2M\right)=1/2$. When, again to approximate the local extrema of the $Q$ distribution, only the top $5\%$ of the partitions is considered, the difference between $Q_3$ and $Q_2$ is not longer zero, but there is wide range of values of $d_1/\left(2M\right)$ and $d_2/\left(2M\right)$ for which $Q_2 > Q_3$ (see Figure~\ref{fig:var_100a}b). This happens when at least one of the merged community is "small", the limit of resolution is related to $\sqrt{M}$~\cite{fortunato07}. Modularity optimization would then tend to aggregate the two groups in one under such circumstances regardless of the other groups' properties. The standard deviation of $Q_3 - Q_2$ in the top $5\%$ behaves differently from what is observed for $q=100$. The maximal standard deviation is obtained for
homogeneous partitions, while it decreases as the partition becomes more and more imbalanced as can be seen in Figure~\ref{fig:var_100a}c. 

\begin{figure*}
\includegraphics[width=0.3\textwidth]{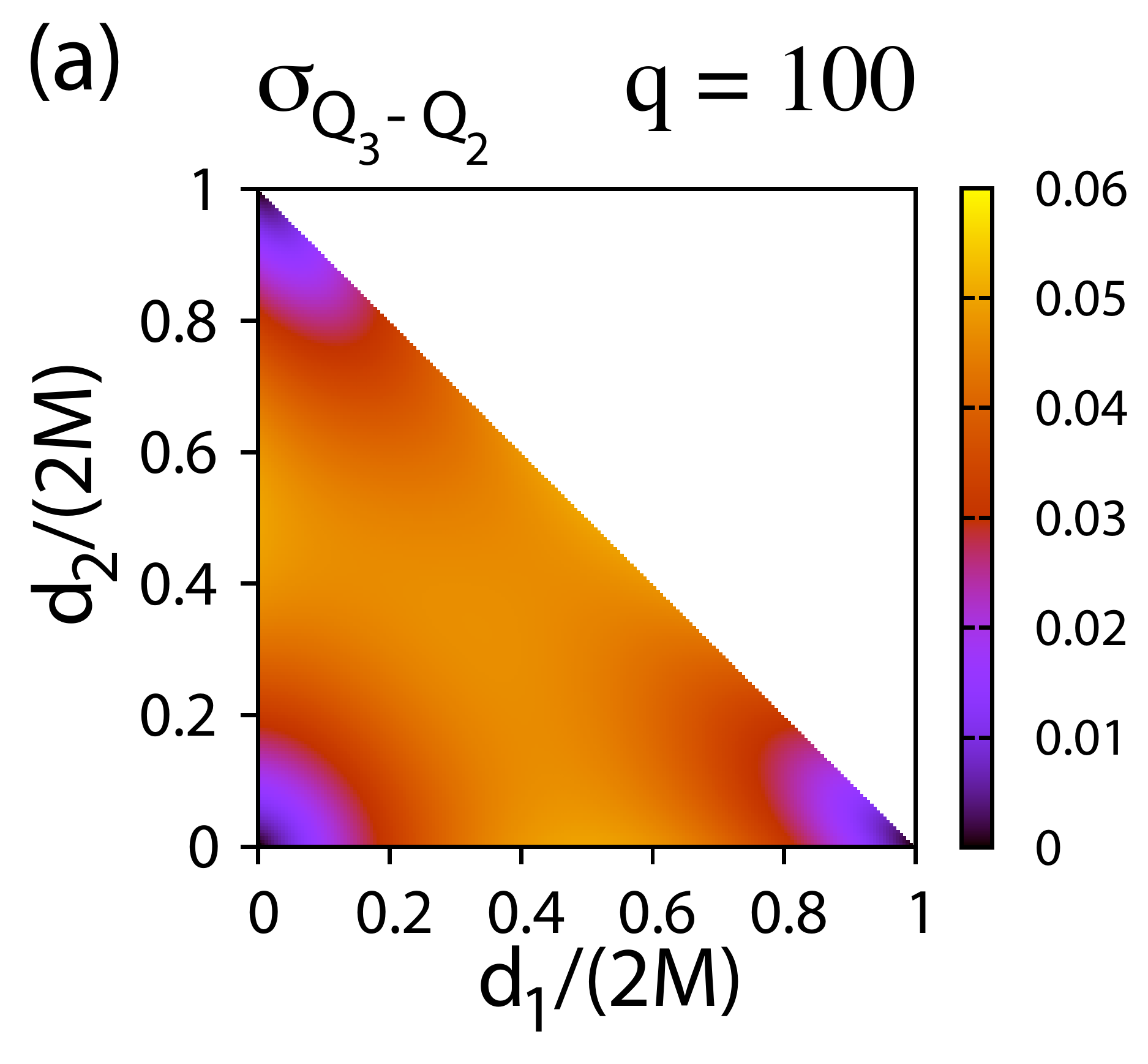}
\quad
\includegraphics[width=0.3\textwidth]{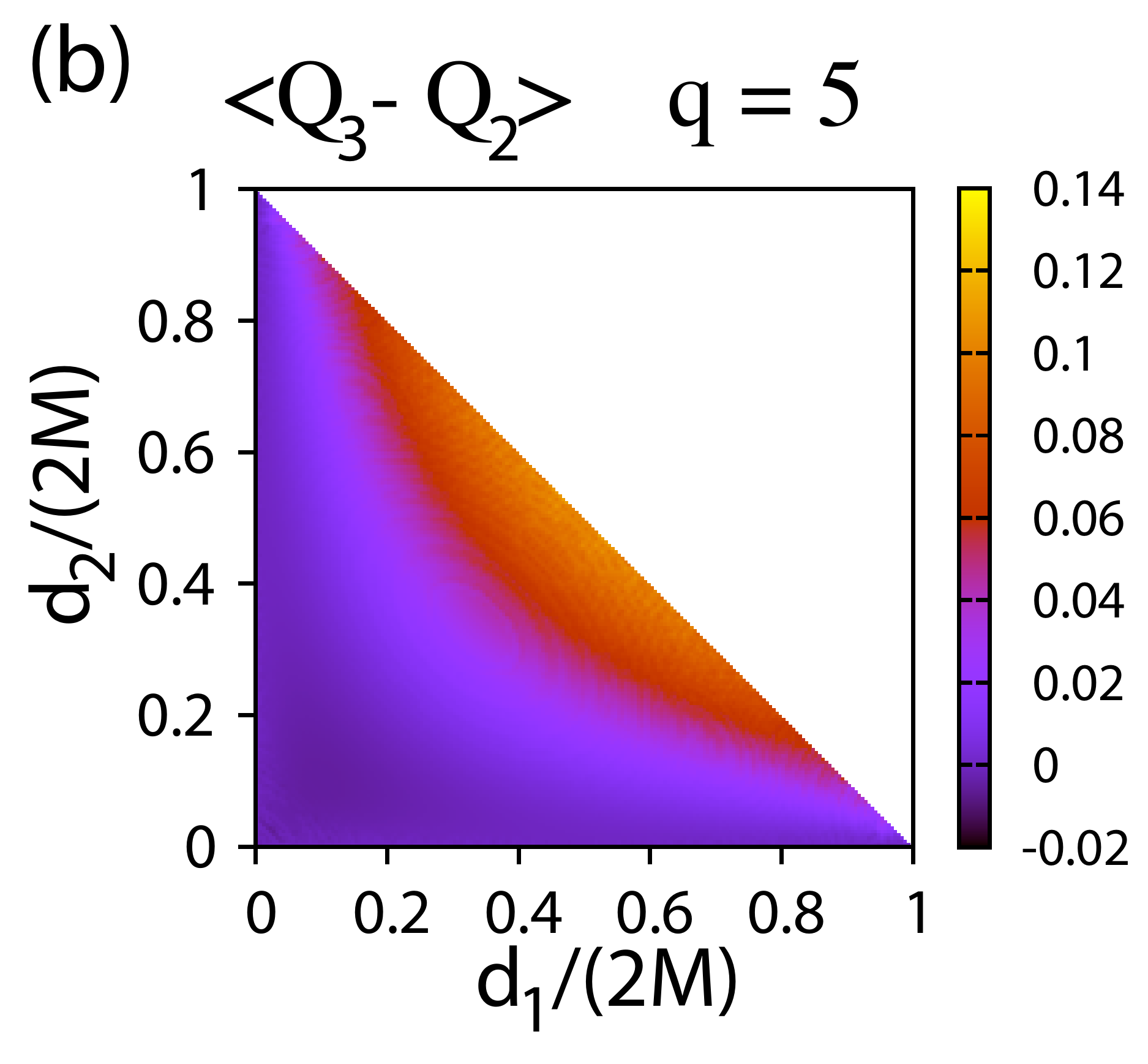}
\quad
\includegraphics[width=0.3\textwidth]{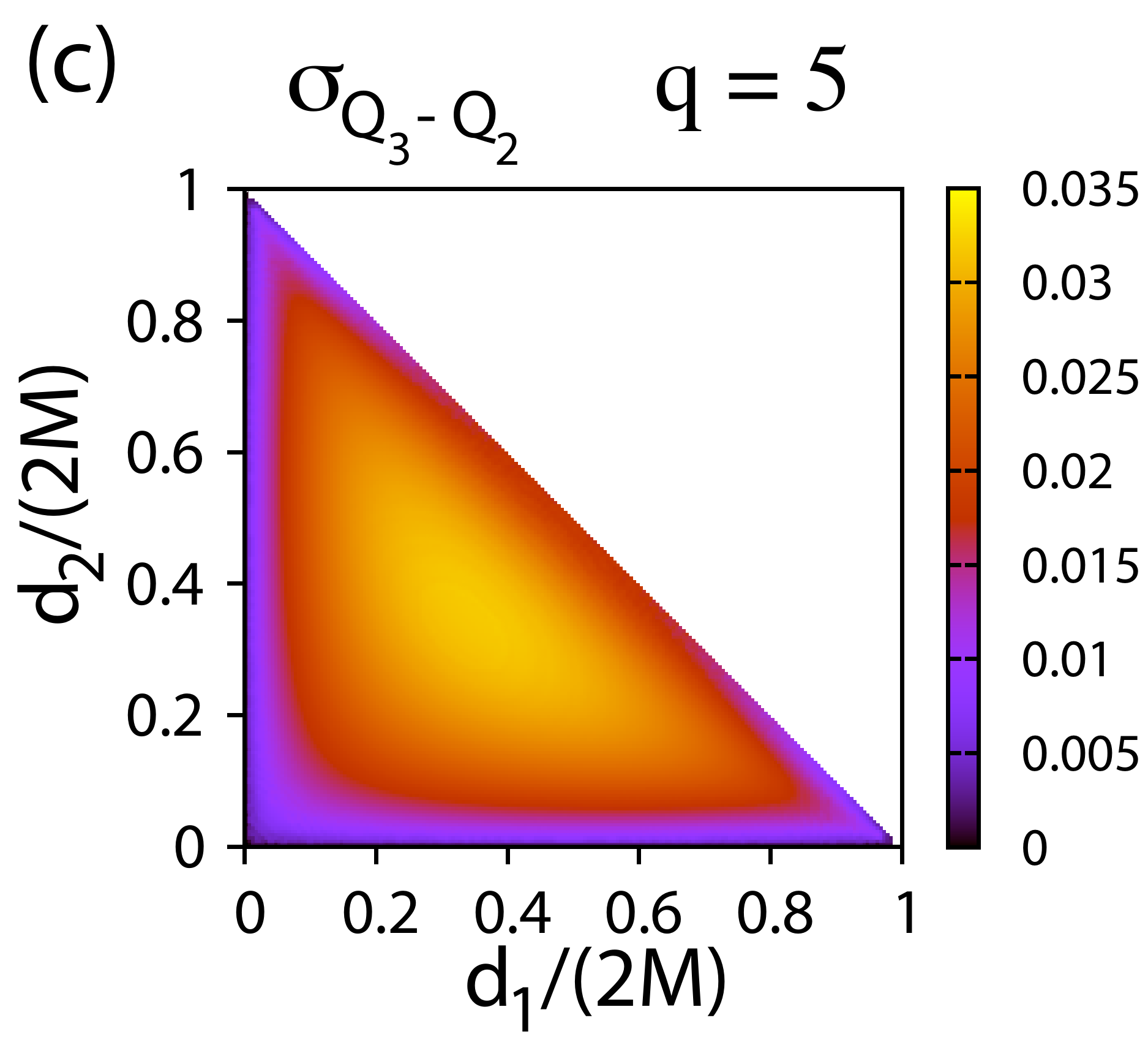}
\caption{(Color online) Fixed the partitions corresponding to the top
$q\%$ of modularity, we compute the average and standard deviation, over this ensemble,
 of the difference $Q_3-Q_2$ as a function of the relative degrees 
of the groups [i.e., $d_1/\left(2M\right)$ and $d_2/\left(2M\right)$]. For $q=100$, the 
average is always zero, while the standard deviation (panel a) grows as 
$d_1/\left(2M\right)$ and $d_2/\left(2M\right)$ tends to $1/2$ (i.e., the third community is empty).
For $q=5$, there is a region in which $Q_2$ is larger than $Q_3$ (the red region in panel b),
while $Q_3-Q_2$ grows as $d_1/\left(2M\right)$ and $d_2/\left(2M\right)$ tends to $1/2$. The standard deviation (panel c) 
is maximal for an homogeneous split of the network [i.e., $d_1/\left(2M\right)=d_2/\left(2M\right)= 1/3$] and
regularly decreases to zero as one move far from the homogeneous split. Here we set $M=100$.}
\label{fig:var_100a}
\end{figure*}

\section{Statistical significance of partitions}
\label{sec:stat} 

The most important application of finding an explicit form for the distribution of the modularity values of the partitions of the random graphs of a null model, as the configurational model, is that the extremes of the distribution offer comparison points to establish the statistical significance of the partitions of equivalent real networks~\cite{guimera04,lancichinetti10}. Given the
degree sequence of the communities $\{d_\alpha\}$, Equation~(\ref{eq:prob_mod}) 
provides the computation of the probability distribution of the modularity function $\mP_C\left(Q\left|\{d_\alpha\}\right.\right)$. In order to consider the different partitions of a graph, we need to obtained the unconditional probability $\mP_C\left(Q\right)$ (only conditioned to the node degree sequence). This probability can be obtained from the convolution  
\begin{equation}
\mP_C\left(Q\right) = \sum_{\{d_\alpha\}} \mP_C\left(Q\left|\{d_\alpha\}\right.\right) \mP_C\left( \{d_\alpha\} \right) \;,
\label{eq:prob_uncond}
\end{equation}
where $\mP_C\left( \{d_\alpha\} \right)$ depends also on the
degree sequence of the nodes in the network (i.e., $\{k_i\}$). 
The computation of this probability is very expensive and we have done it only for $C=2$. In this case, the number of
partitions in which one of the groups has degree $d_1$ can be obtained
as
\begin{equation}
\mG_2\left(\left\{d_1\right\}\right) = \sum_{\{n_k\}} \prod_{k} {N_k \choose n_k}\;,
\label{eq:number_bipart}
\end{equation}
where $N_k$ indicates the number of nodes with degree $k$
present in the network and
$n_k$ the number of vertices with $k$ connections belonging to the group. 
Their sum is subjected to the constraints
\begin{equation}
N = \sum_k n_k \qquad \textrm{and} \qquad  d_1 = \sum_k k n_k \;.
\label{eq:bipart_constr}
\end{equation}
The resulting probability can be calculated as $\mP_2\left(\left\{d_1\right\}\right) = \mG_2\left(\left\{d_1\right\}\right) / 2^N$.

We consider next, as examples, three social networks: the unweighted and weighted version of the Zachary Karate Club~\cite{zachary77} and the friendship 
network between Dolphins~\cite{lusseau03}. In Figure~\ref{fig:mod}, we plot the cumulative distribution of $Q$ for the configurational model graphs obtained with these networks nodes' degree sequences. As the main plot shows, the distribution of $Q$ depends on the original network (that is, on the particular nodes' degree sequence). The inset (a) of the figure shows that the conditional distribution of  $Q$ for different values of $d_1$
(i.e., they have same average value, but
different standard deviation) differs and that the resulting unconditional $\mP_2\left(Q\right)$
strongly depends on the shape of $\mP_2\left(\left\{d_1\right\}\right)$ and therefore on the
degree sequence (see Figure~\ref{fig:mod}b). The modularity calculated for the original bi-partitions of these networks
is high when compared with the typical values observed for the bi-partitions of the equivalent graphs generated by the configurational model. The modularities found for the partitions of the real networks are: $Q_{real}=0.37469$ for the unweighted version
of the Zachary Karate Club, $Q_{real}=0.395959$ for the weighted version of the same network and
$Q_{real}=0.374779$ for the Dolphins social network. In all these three cases, the probability of finding such values among all the partitions of the equivalent configurational-model random graphs is quite low. Still this method to evaluate a partition significance presents a bias. Since all the possible partitions are considered for $\mP_C\left(Q\right)$, even those with low modularity and disconnected groups, the partitions found by a modularity optimization algorithm will tend to be generally dubbed as "unlike". A possible solution, in the spirit of our recent work~\cite{lancichinetti10}, is to restrict the sum in Equation~(\ref{eq:prob_uncond}) to a suitable subset of partitions. An example can be the partitions that are local maxima in the $Q_C$ landscape when the random graphs generated by applying  the configurational model to the given network are analyzed. This, however, involves a systematic search for such maxima that goes beyond the scope of this paper.

\begin{figure}
\includegraphics[width=0.45\textwidth]{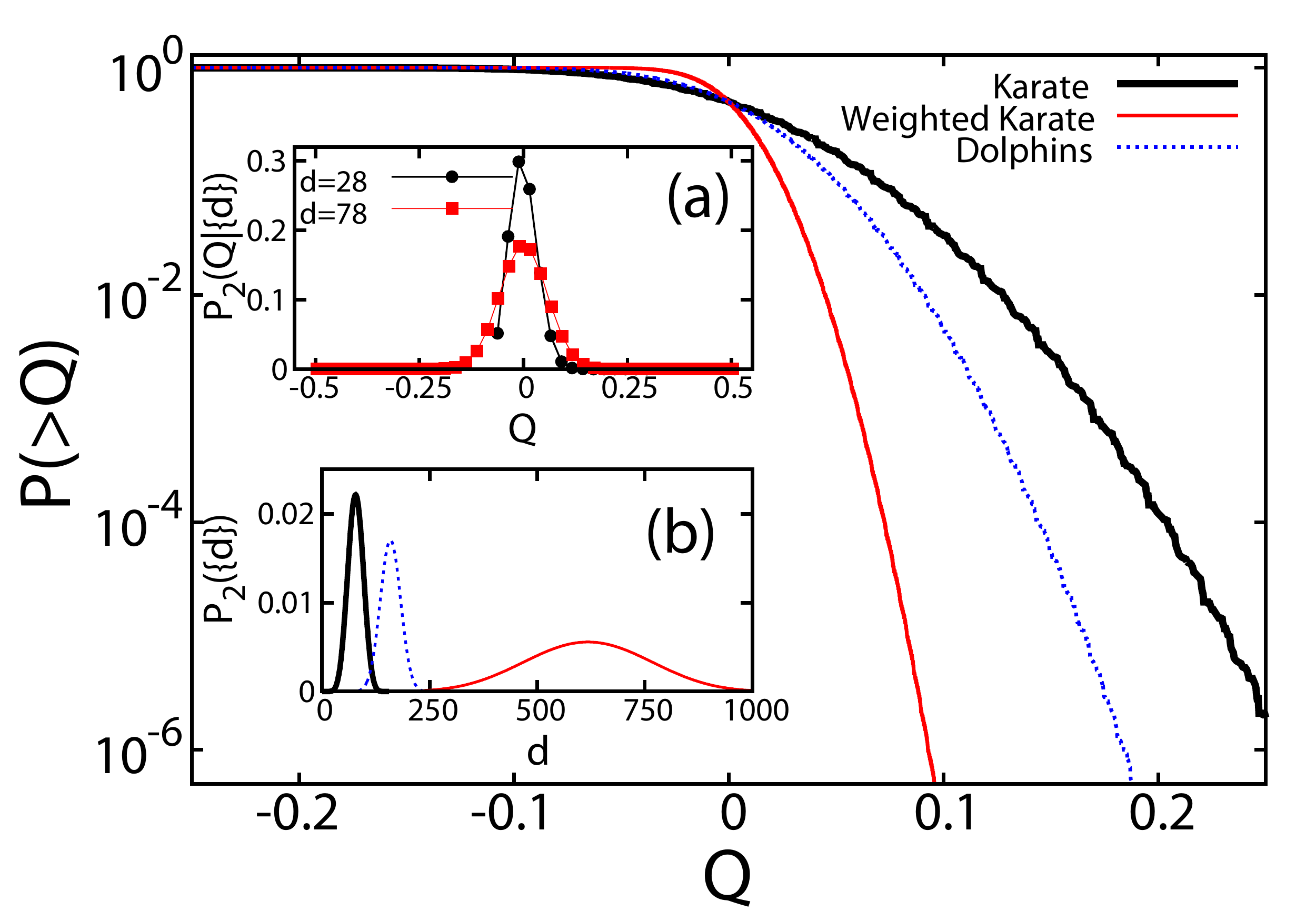}
\caption{(Color online) Cumulative distribution function of modularity for
bipartitions $\mP_2\left(>Q\right)$, calculated for three real networks:
Zachary Karate Club, unweighted (thick black line) and weighted (thin red curve), and
Dolphins social network (dotted blue line). 
In the inset (a), we plot $\mP_2\left(Q\left|\left\{d_1\right\}\right.\right)$, only for the
unweighted version of the Zachary Karate Club, for $d_1=28$ (black circles)
 and $d_1=78$ (red squares).
In the inset (b), we 
report $\mP_2\left(\left\{d_1\right\}\right)=\mG_2\left(\left\{d_1\right\}\right)/2^N$ for the same networks.
}
\label{fig:mod}
\end{figure}

\section{Directed and bipartite networks}
\label{sec:dirntw}

Our combinatorial approach can be easily extended to 
directed and bipartite networks. In these cases, one needs to
distinguish two classes of nodes (bi-partite) or connections (directed).
In the new null model (i.e., the extension of
the configurational model), one needs to reflect this distinction and construct simultaneously two different lists of labels.

We start with the directed networks. Fixing $C$ groups means defining two degree sequences $\{d_\alpha^{in}\}$ and $\{d_\alpha^{out}\}$, corresponding
to the sequences of in-coming and out-going connections, respectively.
In analogy with Eq.~(\ref{eq:constr_2}), each
number appearing in these sequences is represented by the sum of the
in- and out-degrees of all nodes belonging to a given group. 
The total number of possible label sequences that can be formed
is
\begin{align}
\label{eq:tot_nl_dir}
\mT_C^{dir} \left(\{d_\alpha^{in}\}, \{d_\alpha^{out}\} \right) = & \, \\
\frac{M!}{d_1^{in}!\,d_2^{in}!\,\cdots\,d_C^{in}!} & \frac{M!}{d_1^{out}!\,d_2^{out}!\,\cdots\,d_C^{out}!} \nonumber
\,\,\, ,
\end{align}
with constraints given by $\sum_\varphi d^{in}_\varphi=\sum_\varphi d^{out}_\varphi=M$. Eq.~(\ref{eq:tot_nl_dir}) is the product
of the total number of lists of community labels that can be constructed for
 the in-coming and out-going stubs, respectively.
 The total number of lists of community labels
that satisfy the constraints $\{e_{\alpha,\alpha}, e_{\alpha,\beta}\}$ are
\begin{equation}
\mR_C^{dir} \left( \{e_{\alpha,\alpha}, e_{\alpha,\beta}\} \right) =
M!\;\; \prod_{\varphi=1}^{C}\,\prod_{\theta=1}^{C} \frac{1}{e_{\varphi,\theta}!}\;\;,
\label{eq:tot_nc_dir}
\end{equation}
which is the analogous of Eq.~(\ref{eq:tot_nc}), but corrected in this case for
the absence of symmetry (i.e., it may happen that $e_{\varphi, \theta} \neq e_{\theta, \varphi}$). 
The probability to observe a configuration
with intra- and inter-community connectivities
given by $ \{e_{\alpha,\alpha}, e_{\alpha,\beta}\}$ is again the ratio  $\mR_C^{dir}/\mT_C^{dir}$, while
the marginal distribution for the only intra-community connections 
$\{e_{\alpha,\alpha}\}$ can be calculated by summing over all values of the inter-community arcs
subjected to the constraints $d_\varphi^{in} = \sum_\theta e_{\theta, \varphi}$ and $d_\varphi^{out} = \sum_\theta e_{\varphi,\theta}$. 
As in the case of undirected networks, for $C=2$ and $C=3$
no sum is effectively required and the computation of
the marginal probabilities is straightforward. For $C=2$ for example,
we obtain
\begin{align}
\label{eq:prob2dir}
\mP^{dir}_2 \left( \left\{e_{1,1}\right\} \right) = & 
\frac{d_1^{in}!\,\left(M-d_1^{in}\right)!}{M! \; e_{1,1}!\,\left(d_1^{in}-e_{1,1}\right)!} \\
& \times \frac{\,d_1^{out}!\,\left(M-d_1^{out}\right)!}{ \,\left(d_1^{out}-e_{1,1}\right)!\,\left(M-d_1^{in}-d_1^{out}+e_{1,1}\right)!} \,\,, \nonumber
\end{align}
with average $\langle e_{1,1} \rangle = d_1^{in} d_1^{out} / M$.
Eq.~(\ref{eq:prob2dir}) can be used directly for the computation of
the probability distribution of the modularity since, for directed networks, $Q_C$ is defined with an expression similar to the one in Eq.~(\ref{eq:mod}) for undirected networks (only the term for the expected value of internal links in the null model changes)~\cite{leicht08}.

A similar procedure also applies to bipartite networks. In this case
nodes are distinguished in two classes and only vertices belonging
to different classes can be connected. The equations valid for the case
of directed networks can be directly applied to bipartite networks.
There are two different definitions of modularity for bipartite networks.
In the definition of Barber~\cite{barber07}, modules can be constructed
by nodes of both classes and therefore the probability distribution of the modularity can be calculated
directly from the previous equations. The definition of
Guimer\'a {\it et al}.~\cite{guimera07} differently requires that modules are composed only
of vertices of the same type. Our equations need to be modified
and in particular Eqs.~(\ref{eq:tot_nl_dir}) and~(\ref{eq:tot_nc_dir})
should take into account explicitly the presence of $C_1$ and $C_2$ groups
with different type of nodes instead of only $C$ modules.

\section{Summary and Conclusions}
\label{sec:concl}

The study of the community structure of networks 
has attracted much attention during last years.
Most of the work performed in this field of research
has focused on the so-called modularity function,
which has become a standard in this context with widespread usage in many different disciplines. Modularity has the nice characteristics of abstracting into a single number the strength and significance of the whole community structure of a network. Modularity is based on the comparison of the level of internal links in a given graph partition and the expected value of this quantity in the configurational model. This model generates
the ensemble of all uncorrelated networks compatible with the one under study
and therefore constitutes a good term of comparison for the evaluation
of correlations as those at the basis of the existence of communities.
In this paper, we study the modularity via complete enumeration of the partitions of the networks generated by the configurational model.
Our combinatorial approach allows to formulate exact calculations
in the framework of the null model and therefore write
an equation for the probability distribution function of the modularity.
Thanks to this, we are able to study several interesting
features of modularity. We focus on the so-called resolution
limit of modularity, which is statistically
observable in the best partitions of the configurational model, and on the properties of the top ranking instances of the modularity that can be related to the local maxima in the $Q_C$ landscape.
We additionally study an estimator of the statistical significance
of partitions in networks by measuring how probable is the possibility
to observe a particular value of the modularity in the configurational model. Although as warned in the text, this technique is better applied in a distribution of $Q_C$ restricted to a smaller, more selective, set of partitions.

\begin{acknowledgments}
AL and JJR are funded by the EU Commission projects 238597-ICTeCollective and 233847-FET-Dynanets, respectively.
\end{acknowledgments}


\begin{thebibliography}{99}
\bibitem{note} The configurational model may lead to the generation of a network composed of more than one connected component. The probability to observe this event depends on the degree sequence and in general is negligible for sufficiently high values of the average degree.
\end{thebibliography}
\end{document}